%% file: sample-sigconf-biblatex.tex
\documentclass[sigconf,natbib=false,nonacm]{acmart}
\AtBeginDocument{%
  }

\input{packages}
\input{macros}

\setcopyright{none}
\copyrightyear{2018}
\acmYear{2018}
\acmDOI{XXXXXXX.XXXXXXX}

\RequirePackage[
  datamodel=acmdatamodel,
  style=acmnumeric,
  ]{biblatex}

\begin{document}

\title{Probabilistic Programming Meets Automata Theory}
\subtitle{ Exact Inference using Weighted Automata}

\author{Dominik Geißler}
\orcid{0009-0008-8069-1417}
\affiliation{%
  \institution{TU Berlin}
  \city{Berlin}
  \country{Germany}
}

\author{Tobias Winkler}
\orcid{0000-0003-1084-6408}
\affiliation{%
  \institution{RWTH Aachen University}
  \city{Aachen}
  \country{Germany}}

\renewcommand{\shortauthors}{D. Geißler and T. Winkler}

\begin{abstract}
\emph{Probabilistic programs} encode stochastic models as ordinary-looking programs with primitives for sampling numbers from predefined distributions and conditioning. Their applications include, among many others, machine learning \cite{deepprohprogramming} and modeling of autonomous systems \cite{autsystems}. The analysis of probabilistic programs is often \emph{quantitative}---it involves reasoning about numerical properties like probabilities and expectations. A particularly important quantitative property of probabilistic programs is their \emph{posterior distribution}, i.e., the distribution over possible outputs for a given input (or prior) distribution. Computing the posterior distribution exactly is known as \emph{exact inference}. We present our current research using \emph{weighted automata}~\cite{DBLP:reference/hfl/Kuich97}, a generalization of the well-known finite automata, for performing exact inference in a restricted class of discrete probabilistic programs. This is achieved by encoding distributions over program variables---possibly with infinite support---as certain weighted automata.
The semantics of our programming language then corresponds to common automata-theoretic constructions, such as product, concatenation, and others.
\end{abstract}

\maketitle
\section{State of the Art}
Many approaches employ \emph{approximate inference}, often using Monte-Carlo sampling, such as \textsc{Stan} \cite{JSSv076i01}, \textsc{WebPPL} \cite{dippl}, or PMCM \cite{DBLP:journals/corr/WoodMM15}. While these tools can be applied broadly, they only approximate the exact posterior distribution and do not yield exact results.
For \emph{exact inference}, methods like the \emph{weakest pre-expectation calculus} \cite{DBLP:conf/stoc/Kozen83,DBLP:series/mcs/McIverM05}---a probabilistic generalization of Dijkstra’s weakest preconditions---has been adapted for Bayesian inference \cite{DBLP:journals/toplas/OlmedoGJKKM18}. Other automatic approaches use \emph{probability density functions}, such as ($\lambda$)\textsc{Psi} \cite{psi,lambdapsi}, or \emph{weighted model counting}, such as \textsc{Dice} \cite{scaling-exact-inference}, which, however, does not support infinite-support distributions. \textsc{SPPL} \cite{SPPL} uses \emph{sum-product expressions} for representing complex distributions. Closely related are methods based on \emph{probability generating functions} (PGFs; e.g., \cite{redip,generatingfunctionsfor,credip,zaiser}); we view our approach as an automata-theoretic counterpart of the latter.

\section{Probability Generating Automata}
In a recent paper~\cite{10.1007/978-3-032-11176-0_16}, we proposed a novel automata-based exact inference method for a certain class of imperative probabilistic programs.
The key idea is to encode joint probability (sub)distributions as weighted automata, and view each program's semantics as an \textit{automaton transformer}: a function translating an automaton for the input into an automaton for the output distribution.

Formally, the weighted automata we consider are defined over a nonempty subset $\semiring'$ of an $\omega$-continuous semiring $(\semiring, +, \cdot, 0, 1)$, where $\semiring'$ represents the admissible transition weights or labels (e.g.,~\cite{DBLP:reference/hfl/Kuich97}).
A weighted automaton is a tuple $\Raut = (Q, M, I, F)$, where $Q$ is a nonempty finite set of states, $M \in \semiring'^{Q\times Q}$ is the transition matrix, and $I \in \semiring'^{1\times Q}$ and $F \in \semiring'^{Q\times 1}$ are vectors of initial and final weights, respectively.
The semantics of $\Raut$---this is not to be confused with the program semantics (see \Cref{sec:section3})---is defined as $\semanticsaut{\Raut} = IM^*F \in \semiring$, where $M^* = \sum_{i\in\N} M^i$.

We instantiate this abstract framework with the semiring of formal power series with nonnegative real coefficients and the program variables as indeterminates (see \Cref{eq:geom} for an example).
We allow only terms of the form $r$ and $r \cdot X$, where $r \in \Rgez$ and $X$ is a program variable, as transition labels.
We call an automaton $\Raut$ of this form a \emph{probability generating automaton (PGA)} if $\semanticsaut{\Raut}$ is a PGF: a formal power series whose coefficients sum up to not more than $1$.
We stress that this encoding also admits infinite-support distributions, as our PGA may have loops.

The following is an example of a PGA (unlabeled transitions have weight 1):
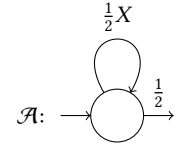
\[
    \begin{tikzpicture}
        \node[state, labeledfinal={{$\frac{1}{2}$}}, initial] (1) {};
        \node[left=0.5cm of 1] (0) {$\Raut$:};
        \draw[loop above] (1) to node[above]{$\frac{1}{2}X$} (1);
    \end{tikzpicture}
\]
First, we can see that this PGA encodes an infinite-support distribution, as it involves a (self) loop containing $X$. The semantics of the automaton is the probability generating function
\begin{align}\label{eq:geom}
    \semanticsaut{\Raut} ~=~ \frac{1}{2}X^0 +\frac{1}{4}X^1 + \frac{1}{8}X^2 + \cdots ~,
\end{align}
which encodes a geometric distribution over $X$ with parameter $\nicefrac{1}{2}$.
We can also think of a PGA as a weighted multi-counter system where variables can only be incremented (but not decremented): every time we take an $X$-transition in a PGA, the value of the program variable $X$ is incremented.
The probability that $X$ has a certain value $n$ is then the sum of the weights of all paths having exactly $n$ many $X$-transitions.  

PGA constitute the semantic domain of our framework.
Similar to other approaches~\cite{psi,generatingfunctionsfor,DBLP:journals/jcss/Kozen81}, our program semantics transforms PGA to reflect the changes the statements have on the underlying probability distribution.
The novelty is that we use common automata-theoretic operations, such as the union, intersection, or concatenation, to represent most of these transformations; see \Cref{eq:semantics} for an example.  

In \cite{10.1007/978-3-032-11176-0_16}, we proved that our PGA transformer semantics is sound with respect to a standard operational small-step semantics based on discrete-time Markov chains.

\section{Automata-Theoretic Inference by Example}
\label{sec:section3}
We illustrate the usage of weighted automata for exact inference with an example. Consider the \emph{goldfish piranha problem}~\cite{chancesbook}: Suppose that you have an opaque bowl with a fish inside---either a piranha ($P=1$) or a goldfish ($P=0$), each with probability $\nicefrac{1}{2}$. You then insert another piranha and, afterwards, catch one of the two fish blindly. Suppose that the fish you caught is a piranha ($R=1$).
\enquote{What,} you wonder while worrying about your fingers, \enquote{is the probability that a piranha (rather than a goldfish) inhabited the bowl initially (i.e., $P=1$)?}

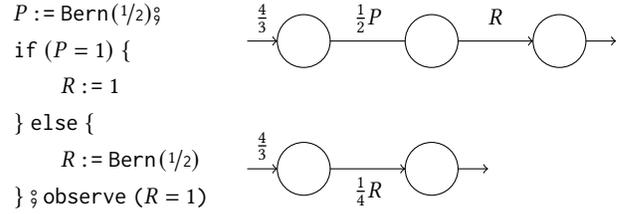
\begin{figure}[t]
    \centering
    \begin{subfigure}{0.15\textwidth}
        
\begin{align*}
    &\assign{P}{\bern{\nicefrac{1}{2}}}\fatsemi\\
    &\texttt{if}~ (P=1) ~\{\\
    &\qquad \assign{R}{1} \\
    &\}~ \texttt{else}~\{ \\
    &\qquad \assign{R}{\bern{\nicefrac{1}{2}}} \\
    &\}\fatsemi\observe{R=1}
\end{align*}
    \end{subfigure}\hfill
    \begin{subfigure}{0.3\textwidth}
        \begin{tikzpicture}
    \node[state, labeledinitial={{$\frac{4}{3}$}}] (0) {};
    \node[state, right of=0] (1) {};
    \node[state, final, right of=1] (2) {};

    \node[state, labeledinitial={{$\frac{4}{3}$}}, below of=0] (3) {};
    \node[state, final, right of =3] (4) {};

    \draw[edge] 
        (0) -- node[above]{$\frac{1}{2}P$} (1)
        (1) -- node[above]{$\vphantom{\frac{1}{2}}R$} (2);
    \draw[edge] (3) -- node[below]{$\frac{1}{4}R$} (4); 
\end{tikzpicture}

    \end{subfigure}
    \caption{The goldfish piranha problem \cite{chancesbook}. \textit{Left}: Probabilistic program encoding the problem. \textit{Right}: Normalized posterior distribution as a (non-connected) PGA. The probability that a piranha was originally in the bowl is $\nicefrac{4}{3}\cdot \nicefrac{1}{2} = \nicefrac{2}{3}$.}
    \label{fig:example-pga}
    \Description{Probabilistic program encoding the goldfish-piranha problem and weighted automaton representing the normalized posterior distribution.}
\end{figure}

In \Cref{fig:example-pga} (left), we can see the goldfish piranha problem encoded as a probabilistic program.
We also depict the PGA resulting from our program semantics (\Cref{fig:example-pga}, right).
It encodes the normalized posterior distribution.
The upper branch represents the outcome where the bowl was originally occupied by a piranha. Multiplying the weights, we obtain the sought-after probability: $\nicefrac 4 3 \cdot \nicefrac 1 2 = \nicefrac{2}{3}$. 

We now showcase how our semantics assembles PGA such as the one in \Cref{fig:example-pga} (right).
We do so by explaining the construction for a representative program instruction: assigning a constant to a variable.
Given a PGA $\Raut$ encoding the prior distribution, the semantics of the assignment $\assign{X}{n}$, where $X$ is a program variable and $n \in\N$, is defined as 
\begin{align}\label{eq:semantics}
    \semantics{\assign{X}{n}}(\Raut) ~=~ \Raut[X/1] \concat \Raut_{\dirac{n}{X}} ~.
\end{align}
We can see that the right hand side of \Cref{eq:semantics} consists of two parts: $\Raut[X/1]$ and $\Raut_{\dirac{n}{X}}$. The former is a \emph{transition label substitution} on $\Raut$, where we replace each occurrence of $X$ by 1. The effect of this operation is to set $X$ to 0.
The second part of \Cref{eq:semantics} involves the PGA for one of the supported distributions, the Dirac distribution with respect to $X$ and the parameter $n$ (whose PGF is simply $X^n$).
We combine the two parts via an automata-theoretic concatenation (or sequential composition), which results in a multiplication of the PGFs encoded by the two automata.
For the sake of illustration, assume that $\Raut$ describes a prior geometric distribution in $X$ with parameter $\nicefrac{1}{2}$ (as in \Cref{eq:geom}) and $n = 1$.
The construction then boils down to the following concatenation (its underlying PGF is stated directly below):
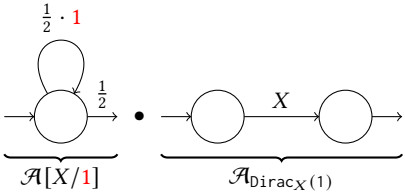
\[
\begin{tikzpicture}
    \node[state, labeledfinal={{$\frac{1}{2}$}}, initial] (0) {};
    \draw[loop above] (0) to node[above]{$\frac{1}{2} \cdot \textcolor{red}{1}$} (0);
    \node[right=0.5cm of 0] (concat) {$\bullet$};
    \node[initial, state, right=0.5cm of concat] (1) {};
    \node[right of=1, state, final] (2){};
    \draw[edge] (1) -- node[above]{$X$} (2);

    \draw [
    thick,
    decoration={
        brace,
        mirror,
        raise=0.5cm
    },
    decorate
] ([xshift=-4mm]0.west) -- ([xshift=4mm]0.east) 
node [pos=0.5,anchor=north,yshift=-0.55cm] {$\Raut[X/\textcolor{red}{1}]$}; 

    \draw [
    thick,
    decoration={
        brace,
        mirror,
        raise=0.5cm
    },
    decorate
] ([xshift=-4mm]1.west) -- ([xshift=4mm]2.east) 
node [pos=0.5,anchor=north,yshift=-0.55cm] {$\Raut_{\dirac{1}{X}}$}; 

\end{tikzpicture}
\]
\begin{align*}
    \semanticsaut{\semantics{\assign{X}{1}}(\Raut)}
    ~=~ \sum_{i=0}^\infty \frac{1}{2^i }\cdot \frac{1}{2} \cdot X 
    ~=~ 2 \cdot \frac{1}{2} \cdot X 
    ~=~ X
\end{align*}
We have similar constructions (indicated in parentheses) for several other programming instructions: coin flips (weighted disjoint union of automata), conditional branching (automata-theoretic products followed by disjoint union), assigning a variable to another variable (transition substitution), observation (product) etc.
To obtain the final PGA for the \emph{normalized} posterior, we reweigh the initial states appropriately in a post-processing step.
See~\cite{10.1007/978-3-032-11176-0_16} for details.

\section{Discussion}
The primary benefit we anticipate from further exploring this approach is the potential to leverage the extensive body of automata theory.

For instance, one of the restrictions of \redip~\cite{redip,credip}, the language we currently support, are \emph{rectangular guards}; that is, Boolean conditions (e.g., in if-statements) can compare variables only to constants, but not to other variables.
We conjecture that we could extend our approach (to a certain degree) with support for non-rectangular guards by using a more sophisticated automaton model (e.g., pushdown automata and their variants).
Ultimately, this would allow us to broaden the class of programs for which inference can be performed exactly.

\section{Ongoing and Future Work}
We are currently developing a prototype implementation.\footnote{The prototype is available online:
\url{https://github.com/dominikgeissler/automata-inference}
}
The current version supports automatic computation of the \emph{unnormalized} posterior distribution, and supports basic PGA minimization.
Adding support for normalization and performing comparisons against other tools for exact inference (e.g., \textsc{Prodigy} \cite{redip,credip} and \textsc{Genfer} \cite{zaiser}) are among the next steps.

Ongoing (unpublished) theoretical research revolves around increasing the expressivity of the supported programming language by incorporating \texttt{while}-statements.
Additionally, we aim to investigate how, for example, certain continuous distributions and non-rectangular guards (see previous section) could be supported.
Another intriguing question is a complete characterization of the expressivity of PGA, that is, to discern the class of distributions they can encode.

\printbibliography

\end{document}

%% file: packages.tex
\usepackage{graphicx}
\usepackage{amsfonts}
\usepackage{amsmath}
\usepackage{stmaryrd}
\usepackage{xspace}
\usepackage{subcaption}
\usepackage{tikz}
\usetikzlibrary{
    arrows,
    decorations.pathmorphing,
    positioning,fit,trees,shapes,shadows,automata,calc,calligraphy,
    patterns,snakes
}
\usepackage{hyperref}
\usepackage{cleveref}
\usepackage{booktabs}
\usepackage{dirtytalk}
\usepackage{mathtools}
\usepackage[table,dvipsnames]{xcolor}
\usepackage{nicefrac}
\usepackage{float}

\usepackage{bbm}

\usepackage{thm-restate}

\usepackage[textsize=scriptsize,disable,bordercolor=white]{todonotes}
\setuptodonotes{fancyline}
\usepackage{csquotes}

%% file: macros.tex
\tikzset{
node distance=1.7cm,
state/.style={draw,circle,outer sep=0pt,minimum size=20pt,inner sep=1pt},
initial/.style={
    initial by arrow, initial text=, initial distance=4mm, draw
},
labeledinitial/.style={
  append after command={
        ([xshift=-4mm]\tikzlastnode.west)
        edge[edge] node[auto]{#1}
        (\tikzlastnode) 
    }
},
labeledinitial top/.style={
    append after command={
        ([yshift=4mm]\tikzlastnode.north)
        edge[edge] node[auto]{#1}
        (\tikzlastnode) 
    }
},
final/.style={
            append after command={
                (\tikzlastnode) edge[edge] ([xshift=4mm]\tikzlastnode.east)
            }
},
labeledfinal/.style={
    append after command={
        (\tikzlastnode) edge[edge] node[auto]{#1} ([xshift=4mm]\tikzlastnode.east)
    }
},
final bot/.style={
state,
            append after command={
                (\tikzlastnode) edge[edge] ([yshift=-4mm]\tikzlastnode.south)
            }
},
labeledfinal bot/.style={
    state,
    append after command={
        (\tikzlastnode) edge[edge] node[auto]{#1} ([yshift=-4mm]\tikzlastnode.south)
    }
},
inv final bot/.style={
state,
            append after command={
                (\tikzlastnode) edge[edge, white] ([yshift=-4mm]\tikzlastnode.south)
            }
},
phantom/.style={draw=none, fill=none},
edge/.style={->},
loop/.style={edge},
loop above/.style={
	edge,
	out=120,
	in=60,
	looseness=8
},
loop below/.style={
    edge,
    out=240,
    in=300,
    looseness=8
  },
  loop left/.style={
    edge,
    out=150,
    in=210,
    looseness=8
  },
  loop right/.style={
    edge,
    out=30,
    in=330,
    looseness=8
  },
underbrace/.style={
decorate,decoration={calligraphic brace,amplitude=10pt,mirror, raise=0.5cm}
},
subautomaton/.style={
draw, rectangle,
text width=1cm,
align=center,
text centered,
inner sep=0.15cm,
minimum height=0.25cm
},
}
\newcommand\sbullet[1][.5]{\mathbin{\vcenter{\hbox{\scalebox{#1}{$\bullet$}}}}}

\newcommand{\concat}{\sbullet}

\newcommand{\Raut}{\mathcal{A}}

\newcommand{\redip}{\texttt{ReDiP}\xspace}

\makeatletter
\providecommand*{\cupdot}{%
  \mathbin{%
    \mathpalette\@cupdot{}%
  }%
}
\newcommand*{\@cupdot}[2]{%
  \ooalign{%
    $\m@th#1\cup$\cr
    \sbox0{$#1\cup$}%
    \dimen@=\ht0 %
    \sbox0{$\m@th#1\cdot$}%
    \advance\dimen@ by -\ht0 %
    \dimen@=.5\dimen@
    \hidewidth\raise\dimen@\box0\hidewidth
  }%
}
\providecommand*{\bigcupdot}{%
  \mathop{%
    \vphantom{\bigcup}%
    \mathpalette\@bigcupdot{}%
  }%
}
\newcommand*{\@bigcupdot}[2]{%
  \ooalign{%
    $\m@th#1\bigcup$\cr
    \sbox0{$#1\bigcup$}%
    \dimen@=\ht0 %
    \advance\dimen@ by -\dp0 %
    \sbox0{\scalebox{2}{$\m@th#1\cdot$}}%
    \advance\dimen@ by -\ht0 %
    \dimen@=.5\dimen@
    \hidewidth\raise\dimen@\box0\hidewidth
  }%
}
\makeatother

\newcommand{\bern}[2]{\ensuremath{\mathtt{Bern}_{#2}\mathtt{(#1)}}\xspace}
\newcommand{\dirac}[2]{\ensuremath{\mathtt{Dirac}_{#2}\mathtt{(#1)}}\xspace}

\newcommand{\semantics}[1]{\ensuremath{\llbracket #1 \rrbracket}}
\newcommand{\N}{\ensuremath{\mathbb{N}}}

\newcommand{\R}{\ensuremath{\mathbb{R}}}
\newcommand{\Rgez}{\ensuremath{\R_{\normalfont\texttt+}}}

\newcommand{\semiring}{\mathbb{S}}

\newcommand{\assign}[2]{\normalfont#1\,\texttt{:=} ~ #2\xspace}

\newcommand{\observe}[1]{\normalfont\texttt{observe} ~ \texttt( #1 \texttt)}

\newcommand{\semanticsaut}[1]{|\!|{#1}|\!|}

\makeatletter
\newcommand{\monus}{\mathbin{\text{\@dotminus}}}

\newcommand{\@dotminus}{%
  \ooalign{\hidewidth\raise1ex\hbox{.}\hidewidth\cr$\m@th-$\cr}%
}
\makeatother